# Cross-Domain Keyword Extraction with Keyness Patterns

Dongmei Zhou[1] and Xuri Tang[2]

1 Wuhan Textile University, 430200, Wuhan, China
2 Huazhong University of Science and Technology, 430074, Wuhan, China
xrtang@hust.edu.cn

**Abstract**

Domain dependence and annotation subjectivity pose challenges for supervised keyword extraction. Based on the premise that second-order keyness patterns exist at the community level, this paper proposes a supervised ranking approach to keyword extraction that ranks keywords with keyness patterns consisting of independent features (such as sublanguage domain and term length) and three categories of dependent features—heuristic features, specificity features, and representavity features. The approach uses two convolutional-neural-network based models to learn keyness patterns from keyword datasets and overcomes annotation subjectivity by training the two models with bootstrap sampling strategy. Experiments demonstrate that the approach not only achieves state-of-the-art performance on ten keyword datasets in general supervised keyword extraction with an average top-10-F-measure of 0.316 , but also robust cross-domain performance with an average top-10-F-measure of 0.346 on four datasets that are excluded in the training process. Such cross-domain robustness is attributed to the fact that community-level keyness patterns are limited in number and temperately independent of language domains, the distinction between independent features and dependent features, and the sampling training strategy that balances excess risk and lack of negative training data.

## 1. Introduction

Keywords (or key-phrases, used interchangeably in this paper) are words and phrases that capture the main topics in a document (Turney 2000). They constitute a bird's view (Nasar et al. 2019) and an extreme summary (Rose et al. 2010; Papagiannopoulou and Tsoumakas 2020) of a document, and they are primary indicators of authors' position and identity (Bondi 2010, 7). The task of keyword extraction plays an important role in Natural Language Processing applications (Rose et al. 2010; Firoozeh et al. 2020) and corpus-based social studies that utilize topic words (Friginal and Hardy 2021) to avoid trivia and insignificant details (Scott and Tribble 2006, 56). Despite increasing attention to this task in years of research (Xie et al. 2023), no approach really emerges as the dominant (Škrlj et al. 2019; Firoozeh et al. 2020). State-of-the-art approaches to keyword extraction generally fall into two categories: unsupervised methods that rely on statistical characteristics and graph representations of texts, and supervised methods that train machine learning models on annotated data.

While state-of-the-art supervised methods offer good enough precision for practical use (Koloski et al. 2021) and perform more effectively than unsupervised methods (Firoozeh et al. 2020; Papagiannopoulou and Tsoumakas 2020), such methods are constrained with two limitations: domain dependence and annotation subjectivity. When choosing supervised approach, one needs to ensure the availability of rich training data, due to keyword heterogeneity in linguistic construction, in input document types and in types required in different applications (Firoozeh et al. 2020). Domain dependence is also attributed to the use of domain-dependent features that are characteristic of the training set (Duari and Bhatnagar 2020).

Supervised methods are also constrained by data subjectivity that is intrinsic and indispensable in the process of assigning keywords to a document (Sterckx et al. 2016; Kim et al. 2013; Papagiannopoulou and Tsoumakas 2020; Sterckx et al. 2018). The behavior of keyword assignment depends on factors such as annotators' previous experience and knowledge and topic background, except for article content (Lu et al. 2020). In addition, keyword annotators may not be able to cover all the good keywords in their assignment process. That is, keywords assigned for a given document are not the only correct ones (Nguyen and Kan 2007), and there are keyword candidates not annotated, or positive unlabelled data (Sterckx et al. 2016). Even within the same language domain, differences exist between the reader- and editor-assigned gold standard (Gallina et al. 2019). Annotation subjectivity leads to data inconsistency, which has large consequences for training and evaluation in the supervised approach (Sterckx et al. 2018).

To overcome the above constraints, this paper posits that there are community-shared second-order keyness patterns learnable from keyword datasets and proposes a supervised approach that extracts keywords by ranking candidate terms with these patterns, hence the name pattern-based keyword extraction, or PKE in short. PKE specifies that the keyness pattern of a candidate term comprises of two independent features, i.e., language domain and term length, and three categories of dependent features — heuristic features, specificity features, and representativity features. To perform the task of keyword extraction, PKE uses a CNN-based (convolutional neural network based) component to learn to identify candidate terms from a document, and then another CNN-based component to learn to rank identified terms. Treating supervised keyword extraction as Positive Unlabelled Learning (Bekker and Davis 2020; Ren et al. 2014; Mordelet and Vert 2014), PKE devices a bootstrap sampling strategy to train the two components. Evaluation of the approach with ten popular keyword datasets yields several significant insights as follows:

- There are community-shared keyness patterns learnable from keyword datasets and supportive of state-of-the-art keyword extraction performance;
- Second-order keyness patterns are temperately language domain independent and are supportive of cross-domain keyword extraction with state-of-the-art performance;
- When training for Positive Unlabelled Learning, there is a sampling size of unlabelled data in a dataset that achieves the best performance in keyword extraction.

## 2. Related works

Recent years have witnessed a number of noteworthy reviews of keyword extraction (e.g. Xie et al. (2023), Nomoto (2023), Papagiannopoulou and Tsoumakas (2020), Firoozeh et al. (2020), and Gupta and Vidyapeeth (2017) etc.). This section only reviews relevant studies, focusing on supervised methods that treat keyword extraction as feature-based binary classification or as learning-to-rank task.

### 2.1 Supervised keyword extraction as binary classification task

This research paradigm uses documents annotated with keywords to train a classifier that determines whether a candidate phase is a keyword nor not (Papagiannopoulou and Tsoumakas 2020). It generally takes two major steps: candidate generation and candidate classification. Candidate generation generally relies on part-of-speech pattern matching and heuristic rules (e.g., Wang and Li (2017)). Candidate classification involves feature selection and classifier choice, which is often the research focus.

Earlier research such as KEA (Witten et al. 1999) and its extensions (Nguyen and Kan 2007; Medelyan et al. 2009) use features such as TF-IDF (term frequency and inverse document frequency), position information, morpho-syntactical characteristics, length, and usage statistics with reference to Wikipedia. The used classifiers include Naïve Bayes and bagged decision trees. Document logical structure in academic publications including components such as titles, authors, affiliations, abstracts, and sections etc. is also used as features for classification (Nguyen and Luong 2010).

Later research expands the set of features and classification algorithms used for the task. Caragea et al. (2014) adds citation contextual features such as in-cited, in-citing, and citation TF-IDF that are based on citation network and uses Naïve Bayes as the classifier. In participating SemEval 2017 Task 10, Wang and Li (2017) use external knowledge that includes English Wikipedia, IEEE taxonomy, and pre-trained GloVe embedding. For classification, Wang et al use random forest for model ensemble of two layers: one layer for output from unsupervised algorithms TextRank (Mihalcea and Tarau 2004) and SGRank (Danesh et al. 2015), and one layer of linear SVM for logistic regression. Florescu and Jin (2018) propose to learn feature representation of phrases by converting a document into an undirected word graph. Duari and Bhatnagar (2020) propose a supervised keyword extractor that models a text as a complex network, and use features such as ngram length, position, centrality, coreness, and clustering coefficient etc. to train a model with Naïve Bayes and XGBoost classifiers.

### 2.2 Supervised keyword extraction as learning-to-rank task

It is argued that keyword extraction is by nature a ranking problem rather than a classification problem (Jiang et al. 2009; Zhang et al. 2017). Treating keyword extraction as a ranking task is to learn a function that assigns keyness scores (the probability of being keywords) to candidate terms on the basis of their features. Jiang et al. (2009) use Ranking SVM on a feature vector consisting of TF-IDF score, phrase length, position information, frequency, and entropy-based uniformity etc. Wang and Li (2011) embed RankBayes in a co-training framework to train a model on features including TF-IDF and position information. Zhang et al. (2017) propose to integrate multidimensional information that includes both features of labelled keywords and structural information of word graph to rank keyword candidates.

Recent literature witnesses the trend of introducing neural networks to the ranking task. Sarkar et al. (2010) train a multi-layer perceptron neural network on a feature set consisting of frequency, IDF, length, and position information of candidates. Mu et al. (2020) first extract span-based feature representation of candidates with a Bi-LSTM on the basis of token features obtained via BERT, and then use a fully connected feed-forward network with sigmoid activation to rank candidates. Xiong et al. (2019) represent candidates with word embeddings, positional embeddings, and visual features such as location, font-size, and HTML DOM features etc., apply a neural network of a convolutional transformer architecture to model the interactions among words inside candidates, and rank candidates with a feed-forward layer. Song et al. (2021) use pre-trained language model RoBERTa to represent words and convolutional neural networks to extract candidate keywords, and then rank the candidates from three perspectives with three modules: a chunking module that measures syntactic accuracy of candidates, a ranking module that measures information saliency between candidates and the input document, and a matching module that measures concept consistency between candidates and the input document. Sun et al. (2021) propose JointKPE that uses pre-trained language models to encode documents, estimates localized informativeness for candidate terms, computes global informativeness for candidates by applying max-pooling upon localized informativeness, and finally ranks candidates based on global informativeness scores.

## 3. Methodology

PKE treats supervised keyword extraction as a learning-to-rank task. This section elaborates on PKE's posit about community-shared keyness patterns, proposes a model to theoretically formalize the pattern-based ranking task, explains the framework that implements the ranking model with convolutional neural networks, and finally expounds the training strategies adopted from Positive Unlabelled Learning.

### 3.1 Pattern-based keyness

The posit of keyness patterns derives from observations in the circle of information retrieval, wherein keywords are index terms (Nomoto 2023) and the concept of keyness is closely associated with the concept of objective aboutness (Maron 1977). While recognizing the subjectivity and user-sensitivity of aboutness (Rondeau 2014; Hjørland 2001), researchers argue that a document has intrinsic topics that are usually agreed upon among different actors in a communication process (Moens 2000, 12) and are constitutional of the basis of tasks like document classification (Beghtol 1986). Objective aboutness exists for a document because language users are able to arrive at a consensus about the topics (that are representable by keywords) of a document.

The above-mentioned objective aboutness is generated from the actual behavior of asking or searching for evidence inside the document concerning candidate terms, and to a certain extent is free from personal and subjective experience (Maron 1977). Such searching activity can be weighing the relative dominance of a candidate in terms of attention, emphasis, space, and frequency devoted to the reference, or achieving unity in the document by following certain rules to select and reject candidates (Wilson 1968). The issue-based theory (Hawke 2018) argues that a topic is identified with a set of distinctions, each consisting of a tuple of concepts that rigidly designate the topic, focusing a discourse on the topic, and allowing others to recede. To recapitulate, the generation of objective aboutness relies on patterns resulted from cognitive analyses involving attention, emphasis, space, frequency, choice, and distinction etc.

The existence of pattern-based objective aboutness allows the possibility of using machine learning tools to simulate the evidence searching behavior. After hypothesizing features that compose keyness patterns, supervised machine learning methods can be used to emulate the searching and ranking behaviours with reference to gold answers in datasets. As gold answers in the datasets are annotated by a number of individuals, the keyness patterns learned by machine learning algorithms will be those shared by the annotator community. Furthermore, when multiple datasets are used in the supervised learning process, the learned keyness patterns will be those generalized across not only individuals but also language domains. As demonstrated by experiments in this study, such community-shared cross-domain keyness patterns are comparatively limited in number so that keyword extraction based on these patterns can achieve satisfactory performance.

### 3.2 Keyness ranking model

Based on the posit mentioned above, we propose the following theoretical keyness ranking model:

$$k_D^t = \tau_D \times \iota_t \times f\left(h_D^t, s_D^t, r_D^t\right) \tag{1}$$

The keyness value $k_D^t$ of a term $t$ in a document $D$ is computed with two feature types[a]: independent features and dependent features, distinguished according to their roles in the ranking process. Independent features are expected to have an impact on dependent features whose values vary along the dimensions specified by independent features. Equation (1) specifies two independent features: sublanguage type of $D$ (denoted by $\tau_D$) and length of $t$ (denoted by $\iota_t$). They have encompassing influence (denoted by multiplication sign ×) on three dependent feature categories: heuristic features ($h_D^t$), specificity features ($s_D^t$), and representativity features ($r_D^t$), all of which are second order features derived from frequency, position, and morpho-syntactical structures of individual terms as detailed in the following sections.

---

[a]This paper use $t$ to denote a candidate term and $D$ to denote a document in question if not otherwise specified.

Distinguishing between independent features and dependent features implies that keyness features should not be used in combination mode. Keyness features are mainly used in combination mode (Firoozeh et al. 2020). It is a common practice for supervised keyword extraction (e.g. Turney (2000), Witten et al. (1999), Jiang et al. (2009), El-Beltagy and Rafea (2010)) to combine a number of keyness features into a vector and use the resulted vector as input to machine learning algorithms. Nevertheless, experiments in this study show that some features (or dependent features) do vary along the dimensions specified by some other features (or independent features) and that taking into account this distinction helps improve keyword extraction performance.

### 3.2.1 Independent features
#### Sublanguage type
A sublanguage type indicates a language domain. It is a consensus that the change of sublanguage type leads to property changes of several keyness features. Aboutness ranking involves the top-down cognitive process that brings expectations of the content and the structure of text types (Beghtol 1986). Richard (1982) reports lexicon diversity in different language domains (Table 1) derived from the examination of about 100 sentences from weather synopses, stock market reports, pharmacology, aviation, recipes, and economics. It can be seen in the table that frequency variation of proper nouns and human nouns among different domains is obvious, intensified by disparity in lexical repetition, synonymy and hyponymy. After experimenting with journal articles, E-mail messages, and webpages, Turney (2000) concludes that a domain-specific algorithm can generate better keywords than general-purpose algorithms. Bouras et al. (2006) examine keyword extraction performance on the datasets of E-mails, academic papers and journal articles, and conclude that a system of keyword extraction needs to be parameterized according to different text types, because term length limit, stop-word lists, and keyword number all vary in different text types.

**Table 1.** Lexicon variation in language domains (adapted from Richard (1982))

| Category | Weather | Stock | pharmacology | aviation | recipes | economics |
|---|---|---|---|---|---|---|
| Proper Nouns | ++ | ++ | A | A | − | A |
| Human Nouns | − | A | A | − | − | A |
| Lexical Repetition | A | + | A | ++ | ++ | + |
| Synonymy | − | ++ | ++ | + | − | - |
| Hyponymy | A | + | ++ | - | + | + |

The symbols indicate different degrees of frequency, as follows: ++ means "far more frequent than in standard English", + "significantly more frequent than in standard English", A "average frequency", - "significantly less frequent than in standard English", − "rare or not used in the sublanguage"

#### Term length
PKE considers term length an independent feature because the length of a term correlates with its occurring positions, morpho-syntactic forms and statistical distribution. Researchers have gradually come to recognize the importance of term length in keyword extraction. Justeson and Katz (1995) demonstrate that the distribution of keywords varies along the length dimension, with bigram keywords dominating in several sublanguage domains. In commenting on the finding that TF-IDF offers very robust performance across different datasets as compared with graph-based approached in Hasan and Ng (2010), Nomoto (2023) observes that many keyword extraction algorithms differ in identifying keywords of different lengths, that setting keywords at the right length is as important as other design choices such as a weighting scheme, and that length is an important but often-dismissed factor of keywords.

### 3.2.2 Dependent features

The three dependent feature categories listed in Table 2 are assembled from earlier literature on both supervised and unsupervised keyword extraction. Each feature category includes several metrics. Details follow.

**Table 2.** Dependent features and their corresponding metrics

| Category | Metrics |
|---|---|
| Heuristic features | casing score; first-occurrence position score; term frequency score; term context diversity score |
| Specificity features | TF-IDF score; effect-size specificity score; lexical specificity score |
| Representativity features | Dispersion: sentence dispersion score |
| | Personalized metrics of centrality: position-rank score; TF-IDF-rank score; lexical-rank score; single-rank score; topic-rank score<br>Topic-based metrics of centrality: eigenvector centrality score; closeness centrality score; betweenness centrality score |

#### Heuristic features

Heuristic features are derived from empirical observation and proven useful for keyword extraction in earlier research. As listed in Table 2, PKE includes four metrics as heuristic features: casing score, first-occurrence position, term frequency score, and term context diversity score.

Casing is important in keyword extraction because many keywords are upper-cased technical terms (Campos et al. 2020). In addition, technical terms are capable of forming new keywords by combining with lower-cased words, such as *British beer import*. PKE computes the casing score of a term $t$ in $D$, as the average upper-case percentage of all the instances of $t$ in $D$, as follows:

$$h^t_{case} = \frac{1}{n} \sum_{t_i}^{n} \frac{\text{Number of upper-cased words in } t_i}{\text{Number of words in } t_i} \tag{2}$$

The position in which a term occurs in a document constitutes an important keyness indicator (Witten et al. 1999; Florescu and Caragea 2017; Campos et al. 2020; Miah et al. 2022). Terms occur earlier in a document and there are more likely to be topics (Campos et al. 2020). After examining five datasets, Miah et al. (2022) report that 51.98% of keywords occur in the 1st region and 7.90% in the 2nd region of an article. PKE computes the first-occurrence position score of $t$ in $D$ as follows:

$$h^t_{pstn} = -ln \frac{\text{Index of the sentence of ts´ first occurrence}}{\text{Nummber of sentences in } D} \tag{3}$$

Frequency is another important keyness indicator, but their relationship is not proportional (Campos et al. 2020). As the length of a candidate term greatly affects its frequency inside a document, PKE distinguishes unigram, bigram, trigram, and quadgram and uses the same method proposed in (Campos et al. 2020) to compute frequency score, as follows:

$$h^t_{freq} = \frac{n_t}{\bar{n}_{ngram} + \sigma_{ngram}} \tag{4}$$

wherein $n_t$ is the frequency of $t$, $\bar{n}_{ngram}$ is the average frequency of the corresponding ngram category (unigram, bigram, trigrams, or quadgrams) of $t$, and $\sigma_{ngram}$ is the standard deviation of the ngram category.

Term context diversity is named term relatedness to context in Campos et al. (2020). The intuitive idea is that a term with more diverse context is less likely to be a keyword. PKE adopts the same calculating method in Campos et al. (2020), who state that this feature is comparatively important.

*Specificity features*

Manual keyword assignment follows specificity as a basic principle. Keywords should be as specific as possible to represent the content of a document and leave out common terms found in other similar documents (Firoozeh et al. 2020). Chuang et al. (2012) uses topic specificity to measure how much a word is shared across topics. PKE uses three metrics to measure how one term contributes to the distinctiveness of the document as compared to other documents, i.e., TF-IDF score, effect-size specificity score, and lexical specificity score.

TF-IDF is used in keyword extraction in various forms (Salton et al. 1975; Witten et al. 1999; Firoozeh et al. 2020; Haddoud and Abdeddaïm 2014; Caragea et al. 2014; Papagiannopoulou and Tsoumakas 2020). El-Beltagy and Rafea (2010) note that the length of a term greatly affects the performance of TF-IDF in keyword extraction. PKE proposes to reduce the effect of high frequency of unigrams by computing term frequency with Equation (4). Therefore, the TF-IDF score is computed as follows:

$$s_{TF\text{-}IDF} = \frac{n_t}{\bar{n}_{ngram} + \bar{\sigma}_{ngram}} \times ln \frac{|D|}{|j : t_i \in d_j|} \tag{5}$$

Effect-size specificity (Scott and Tribble 2006; Scott 1997 2001; Gabrielatos 2018; Gries 2021; Rayson and Potts 2020) characterizes keyness by comparing term frequency in a given text with normalized frequency computed from a reference corpus. It is established by effect-size metrics computed on the following statistics: the length of the document $D$, the size of reference corpus, the frequency of $t$ in $D$, and the frequency of $t$ in reference corpus (Gabrielatos 2018). PKE adopts the measure based on Kullback-Leibler (KL) Divergence in Gries (2020), given below:

$$s_{effect\text{-}size} = p_t^D \times (lnp_t^D - lnp_t^R) \tag{6}$$

wherein $p_t^D$ is the probability of $t$ in $D$, and $p_t^R$ is the probability of t in reference corpus. KL Divergence is well-grounded in information theory and much less correlated with frequency.

Lexical specificity (Ushio et al. 2021) is defined to be the negative logarithm of hypergeometric distribution, as follows:

$$s_{lex\text{-}spec} = -log_{10} \sum_{k=f}^{F} P_{hypergeo}(x = K, N_D, N_R, f_D, f_F) \tag{7}$$

wherein $P_{hypergeo}$ is the probability of $t$ occurring exactly $k$ times in $D$, which is computed according to the hypergeometric distribution parameterized with the number of words in the document ($N_D$), the frequency of the word in the document ($f_D$), the frequency of the word in reference corpus ($f_R$), and the number of words in the reference corpus ($N_R$). As hypergeometric distribution represents the discrete probability of $k$ successes in $n$ draws without replacement, a lower probability of a term occurring with frequency $k$ in the document implies a higher specificity score.

*Representativity features*

Representativity refers to keywords' capability of epitomizing the content of a document by representing the key elements of the document by contrast with other terms that reflect minor aspects (Firoozeh et al. 2020). PKE measures representativity with two sub-types: dispersion and centrality, as listed in Table 2.

Dispersion refers to the degree of evenness to which the occurrence of a word are distributed throughout a document (Gries 2020; Egbert and Biber 2019). PKE adopts the metric proposed in Campos et al. (2020), given below:

$$s_{sentence\text{-}dispersion} = \frac{C_t}{N_D} \tag{8}$$

wherein $C_t$ is the number of sentences $t$ occurs in, and $N_D$ is the number of sentences in $D$.

Centrality is mainly captured with graph-based methods in keyword extraction (Mihalcea and Tarau 2004; Danesh et al. 2015; Ushio et al. 2021; Florescu and Caragea 2017). PKE distinguishes two categories of centrality metrics: metrics based on personalized page-rank (personalized metrics in short) and topic-based metrics (topic metrics in short). Personalized metrics are computed with a graph built with candidate terms as vertices and co-occurrence inside a sentence as the criterion to establish connections. The vertices are then personalized respectively with values of first-occurrence position score, TF-IDF score, lexical specificity, or the default value 1, and are ranked with PageRank algorithm (Brin and Page 1998) to obtain position-rank scores, TF-IDF-rank scores, lexical-rank scores, and single-rank scores.

Topic-based metrics are based on the idea of TopicRank Bougouin et al. (2013). To build a graph from a document, the terms are first clustered to groups, which are used as vertices in the graph, and the connections between vertices are established by sentence co-occurrence of the terms in the groups.PKE uses three topic-based centrality scores: eigenvector centrality score, closeness centrality score, and betweenness centrality score, based on the rationales introduced in Borgatti and Everett (2006). According to their study, both eigenvector centrality and closeness centrality fall into the category of radial centrality, while betweenness centrality belongs to the category of medial centrality. Radial centrality measures coreness that makes sense inside a network with one center with the core-periphery assumption (Borgatti and Everett 2006). Therefore, closeness centrality and eigenvector centrality are good measures of keywords inside documents that are coherent with one topic. For instance, Sarracén and Rosso (2023) use eigenvector centrality for keyword extraction on tweets that are generally small texts and achieve better performance than YAKE. Medial centrality such as betweenness centrality is based on the number of walks passing through a given vertex in a graph. It measures how often a vertex lies on the shortest paths between two vertices inside a graph. As medial centrality does not make the core-periphery assumption and correctly assign particularly high centrality scores to nodes serving as bridges between subgroups inside a graph (Borgatti and Everett 2006), such centrality scores should be better at identifying keywords inside documents consisting of multiple subtopics. Load centrality (Goh et al. 2001), which is equivalent to betweenness centrality that quantifies which nodes are important in the transport dynamics inside a graph, is used for keyword extraction in Škrlj et al. (2019).

### 3.3 Implementation framework

Figure 1 gives the overall framework with which PKE implements the posit of pattern-based keyword extraction and the keyness ranking model of Equation (1). The framework processes an input document with three sequential components: pre-processing, candidate generation, and candidate ranking, and output ranked term groups for further process. The pre-processing component normalizes input documents and parses them with a dependency parser. The component of candidate generation generates ngrams from input documents, identifies candidate terms inside the generated ngrams with a convolutional-neural-network (CNN) based model, and clusters these candidate terms into groups. The candidate-ranking component ranks the terms by generating keyness feature patterns for candidate terms and ranking with another CNN based model. The TF-IDF data are also collected from the same datasets used for training the two neural networks. Details are given below.

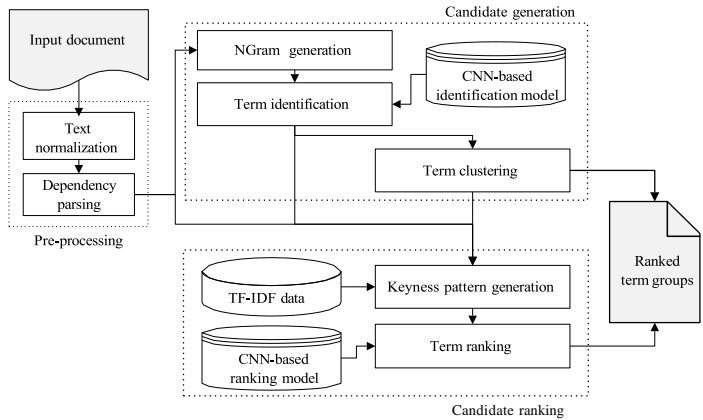

**Figure 1.** Framework of pattern-based keyword extraction.

### 3.3.1 Pre-processing

Two operations are applied to an input document in the pre-processing component. First, the document is normalized by replacing hashtags, URLs, bullet points with tags, fancy quotation marks with ASCII equivalents, and by deleting redundant hyphens and white spaces. Second, a dependency parser is applied to the document to obtain morpho-syntactic information including part-of-speech, casing, dependency type, and being-a-stopword-or-not etc., which is used in candidate generation and ranking process.

### 3.3.2 Candidate generation

The component of candidate generation uses three steps to generate candidate terms. The first is to generate ngrams ($1 \leq n \leq 4$) from the input document. The second is to identify candidate terms among the generated ngrams according to the property of well-formedness, i.e., being syntactically and semantically well-formed (Firoozeh et al. 2020). Phrases such as *number of, like the, did you mean will* not be identified as candidate terms because they are not syntactically and semantically autonomous and are requisite of further information to make sense. PKE models this property with the convolutional networks given in Figure 2, which take as input a vector of features and output two probabilities: ill-formed probability (*I-prob*) and well-formed probability (*W-prob*). A ngram is identified a candidate term by computing the final well-formedness score $\omega = W\text{-}prob - I\text{-}prob$. It is a candidate term if $\omega > 0$.

In Figure 2, the input vector of a ngram comprises of features of each word inside the ngram, which is a quadruple defined below:

<part-of-speech, case-status, is-stop-word, dependency-type>

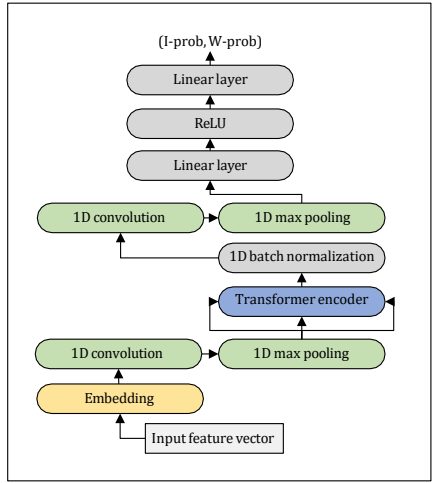

**Figure 2.** Candidate identification model , outputting (ill-formed probability ($I\_prob$), well-formed probability ($W\_prob$)).

The quadruple contains four pieces of information: the part of speech of the word, the word being in upper case or lower case, the word being a stop word or not, and the dependency relation in which the word serves as the dependent to its head. For example, the input feature vector of the trigram *Atlantic hurricane season* is as follows:

*[<NNP, UPPER-CASE, NOT-STOP, amod>,*
*<NN, LOWER-CASE, NOT-STOP, compound>,*
*<NN, LOWER-CASE, NOT-STOP, nsubj>,*
*<UNK, UNK, UNK, UNK>]*

wherein *NNP* and *NN* are parts of speech, *amod*, *comound*, and *nsubj* are dependency types, and *UNK* is the padding tag.

The neural networks in Figure 2 have three major parts. The first is the embedding layer that transforms category names of parts of speech, case status (*UPPER-CASE* and *LOWER-CASE*), stopword status (*IS-STOP* and *NON-STOP*), and dependency relation types into embeddings. The second part consists of two convolution layers, two max pooling layers, and one transformer encoder layer, thus constructed to capture morpho-syntactic patterns. The third part consists of two linear layers and one non-linear activation layer (ReLU), used to map the morpho-syntactic patterns to binary probability.

The third step in this component is term clustering, used to address the keyword property of citationness (Firoozeh et al. 2020). As metadata of a document, keywords are often lemmatized to meet the need of citation. In addition, the topic that a keyword represents may also occur in a document in synonymous forms (or semantically-equivalent forms (Kim et al. 2013)) except for those inflective forms, as illustrated in Table 3. Ignoring keyword citationness leads to underestimation of actual system performance, because many semantically equivalent keywords are not counted as correct (Kim et al. 2013). Clustering candidate terms will yield term groups, or topics to be used in the ranking process.

### 3.3.3 Candidate Ranking

The candidate ranking component takes two steps to rank candidate terms: keyness pattern generation and term ranking. In the first step, a total of nineteen features are collected to form a keyness pattern for each candidate term, including two independent features (ref. Section 3.2.1), sixteen dependent features (ref. Table 2), plus the well-formedness score (ref. Section 3.3.2). When computing metric scores for dependent features, information is gathered from multiple sources including the dependency-parsed document, term clustering, and TF-IDF data gathered from keyword datasets.

**Table 3.** Examples of citationness taken from the dataset DUC (Wan and Xiao 2008)

| Topic | Synonymous forms |
| --- | --- |
| emergency landing | landing, crash landing, landing Wednesday, emergency landing |
| tail engine tail | tail engine, huge tail, tail section |
| flight crew | crew, crew member, flight crew |

The second step, i.e., term ranking, is performed by the neural networks given in Figure 3. The impact of independent features upon dependent features as specified in Equation (1) is implemented in Figure 3 with a dot product (indicated by ⊗) of embedded sublanguage type and term length and the feature vector composed of the rest seventeen term scores. The resulted vector is further encoded with positional information using the position scheme proposed in Vaswani et al. (2017). Two transform encoders, three convolutional layers, and three pooling layers are used to capture keyness patterns and the last linear layer output a tuple consisting the negative keyness score (*N-score*) and positive keyness score *P-score*), from which the final keyness score is computed as follows: $r = P\text{-}score - N\text{-}score$.

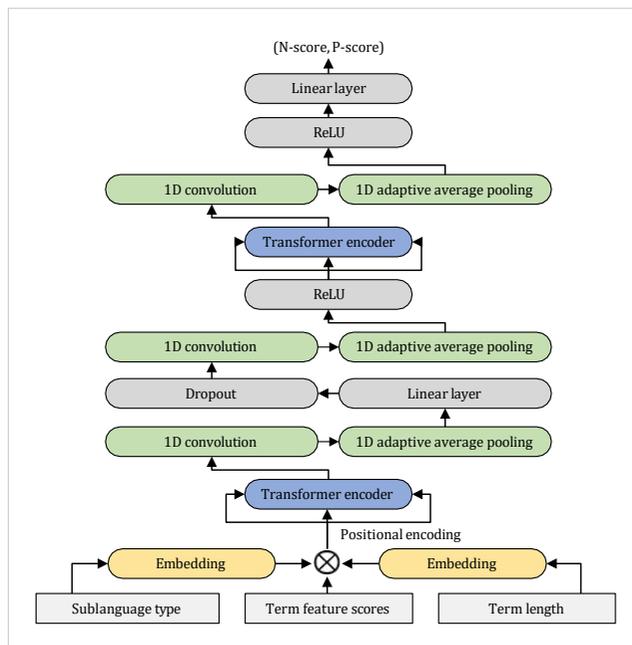

**Figure 3.** Candidate ranking model, outputting negative keyness score (N-score) and positive keyness score (P-score).

### 3.3.4 Ranked term groups

Instead of outputting terms with their ranking scores, PKE outputs the term groups obtained by Term Clustering in the Candidate generation component that are ranked according to the highest ranking score inside each term group. Therefore, PKE outputs should be treated as ranked topics but not ranked terms. Accordingly, the evaluation of PKE's performance is the evaluation of the system's capacity in ranking topics of a document rather than terms. This is in line with the concept of keywords as the meta-data of a document.

### 3.4 Model training and Positive Unlabelled Learning

As in other supervised methods, the training of the two neural networks in Figure (2-3) in PKE is confronted with the subjectivity intrinsic in keyword datasets. The phenomenon that training data contain some annotated data and a lot of unlabelled data is often addressed within the machine-learning branch termed Positive Unlabelled Learning (Bekker and Davis 2020; Ren et al. 2014; Mordelet and Vert 2014). More specifically, supervised keyword extraction can be treated as a task of positive unlabelled (PU) learning with separability, wherein keywords can be distinguished from non-keywords although individual annotators may disagree with each other on which candidate term qualifies as a keyword and sometimes it is difficult to decide whether a term is a keyword. This inductive PU learning carries the following goal:

$$f : k_D^t \to R \qquad (9)$$

i.e., to learn a function $f$ that predicts the keyness score $k_D^t$ of $t$ in $D$ that can be any real number in $R$ by training o na positively labelled term set and an unlabelled term set.

Like other machine learning methods, neural networks are also based on counts of positive and negative examples in training data (Bekker and Davis 2020). It is not appropriate to indiscriminately input all data for training, because the count of unlabelled positive instances might be equal or even larger than the count of labelled instances. Mordelet and Vert (2014) argue that it can be useful to train such models by penalizing more false negative labels than false positive labels inside datasets, given the fact that positive instances are known to be positive while unlabelled examples contain hidden positive ones. They propose to perturb the original training set through bootstrap re-sampling, and specifies the upper bound of excess risk of discriminating P (labelled dataset with size $p$) from U (a sub-sample dataset with size $k$) with contamination rate $\hat{\gamma} = \frac{C_+}{k}$ (the proportion of positive examples $C_+$ in U) to be computed with Equation (10) below:

$$Risk \leq a \times \frac{p^{-1/2} + k^{-1/2}}{1 - \hat{\gamma}} \qquad (10)$$

Assuming $k = p \times \vartheta$, i.e., the size of U is measured as the ratio $\vartheta$ of $p$, Formula (10) can be transformed into Equation (11) below:

$$Risk \leq a \times \frac{p^{-1/2} + k^{-1/2}}{1 - \hat{\gamma}} = a \times \frac{p^{-1/2}(1 + \vartheta^{-1/2})}{1 - \hat{\gamma}} \qquad (11)$$

As p and $\vartheta$ are fixed in a given dataset, it can be inferred from Equation (11) that the higher $\vartheta$ is (or the more are sampled), the higher the excess risk. On the other hand, a smaller $\vartheta$ means that fewer negative training data participating in the training process, which results in insufficient training and poorer performance. Therefore, it can be inferred that there exists a $\vartheta$ that balances both excess risk and the number of negative training data so that the best performance can be achieved.

When several datasets with different contamination rates are used for training, it will be more difficult to decide on an appropriate $\vartheta$ value. In Equation (11), the contamination rate $\hat{\gamma}$ is negatively related to $\vartheta$. That is, a lower annotation agreement of a dataset leads to a higher contamination rate so that more positive unlabelled data and fewer true negative data should occur in U. In such cases, a higher $\vartheta$ is required so that sufficient negative training data can be included in the training data.

Accordingly, PKE follows the procedure below to train the neural networks in Figure 2:

(1) Initialization: $M_{epoch}$ is the num of training epochs and $\varepsilon$ is the keyness filter;
(2) For each epoch $i$, construct the training dataset $D_{train}$ as follows:

    a. If $i = 1$, process each dataset by (i) dividing input dataset into labelled set $P_{key}$ with $p$ instances and unlabelled set U, (ii) obtaining $U_i$ by sampling $p$ instances from U, and (iii) adding both $U_i$ and $P_{key}$ to $D_{train}$;
    b. If $i > 1$ and $i$ mod 5 is not 0, use the same $D_{train}$ as the last epoch;
    c. If $i > 1$ and $i$ mod 5 is 0, iteratively process each dataset by (i) dividing the input dataset into labelled set $P_{key}$ with $p$ instances and unlabelled set U, (ii) obtaining $U_i$ by sampling $p$ instances from U, (iii) using the trained model M to predict $U_i$, (iv) filtering out those instances in $U_i$ if $M(\,\cdot\,) > \varepsilon$ , and (v) adding the resulted $U_i$ and $P_{key}$ to $D_{train}$;

(3) Start training with $D_{train}$, obtain the identification model M.

PKE trains the candidate ranking model in Figure 3 with the following procedure:

(1) Initialization: $\vartheta$ is the sampling ratio as defined in Equation 11;
(2) For each epoch, process each dataset as follows to obtain $D_{train}$: divide the input dataset into keyword set $P_{key}$ (with $p$ instances) and non-keyword set U, randomly select $p \times \vartheta$ number of instances from U, combine the selected terms with $P_{key}$, and add the resulted combination to $D_{train}$;
(3) Start training with $D_{tran}$ and go back to step (2) to modify $D_{train}$ for every five epochs.

## 4. Experiments and Discussions

This section reports the experiments conducted to evaluate both candidate identification and candidate ranking in PKE, focusing on the comparison between PKE and state-of-the-art keyword extraction systems, and on PKE's competence in cross-domain keyword extraction.

### 4.1 Experiment Setup

#### 4.1.1 Datasets

This study selects ten publicly available keyword datasets of English language (listed in Table 4) for evaluation, taking into account factors such as variation in sublanguage domain, term length, document length, number of keywords per document, and document number. The ten datasets fall into five sublanguage domains: miscellaneous (a mixture of different domains), agriculture, computer science, general science, and medical science. These five sublanguage domains are not strictly distinguished, some (such as general science) overlapping with others (such as agriculture and computer science). The average length of terms varies from 1.26 in CiteU-Like to 2.16 in SemEval-2010. The diversity in document number is also noteworthy as PKE is a supervised method and its performance is expected to rely on training data size. Also note that PKE is only concerned with keywords available in documents and the keywords absent from documents are ignored.

Annotation agreement is reported for four of the ten datasets. The kappa statistics for inter-agreement among annotators in DUC is 0.70 (Wan and Xiao 2008). KP-Crowdsource is annotated via crowdsourcing with the average agreement between annotators to be 55%. CiteU-Like is tagged by 332 indexers with a consistency score of 37.7 (Medelyan et al. 2009). For Wiki20, the average consistency among the teams is 30.5%, indicating that the teams generally do not have mutual agreement on what constitutes a topic for a document (Medelyan et al. 2010).

For evaluation purpose, each of the ten datasets is split into a training dataset and a test dataset in an 8:2 ratio. The training datasets are used in training both the identification model and the ranking model, while the test datasets are used to evaluate the performance of the two models.

**Table 4.** Keyword Extraction Dataset

| Name | Source | Doc. type | Sub-language | Doc. num. | Avg. word num. | Avg. keyword num. | Avg. keyword length | Avg. abs. per. |
|------|--------|-----------|--------------|-----------|----------------|-------------------|---------------------|----------------|
| KP-Crowd-Source | Marujo et al. (2012) | news | misc. | 500 | 506 | 49.21 | 1.39 | 8.19 |
| KP-Times | Gallina et al. (2019) | news | misc. | 280,000 | 921 | 5.20 | 1.76 | 54.16 |
| DUC | Wan and Xiao (2008) | news | misc. | 308 | 884 | 8.06 | 2.08 | 6.31 |
| FAO780 | Medelyan and Witten (2008) | article | agricul-ture | 780 | 32,262 | 7.98 | 1.62 | 26.92 |
| FAO30 | Medelyan and Witten (2008) | article | agricul-ture | 30 | 22,911 | 32.33 | 1.61 | 32.47 |
| Wiki20 | Medelyan et al. (2010) | article | computer science | 20 | 7,715 | 37.30 | 1.7 | 55.77 |
| CiteU-Like | Medelyan et al. (2009) | paper | misc. | 180 | 8,622 | 17.42 | 1.26 | 35.55 |
| NYN | Nguyen and Kan (2007) | paper | science | 210 | 8,692 | 11.45 | 2.04 | 15.23 |
| PubMed | Aronson et al. (2000) | paper | medicine | 500 | 5,482 | 16.36 | 1.61 | 59.63 |
| SemEval-2010 | Kim et al. (2013) | paper | misc. | 243 | 9,646 | 15.48 | 2.16 | 13.34 |

Doc. stands for document, num. for number, avg. stands for average, misc. for miscellaneous, abs. per. for absent percentage. Absent keywords are keywords that cannot be strictly matched without stemming or lemmatization.

### 4.1.2 Experiment procedure and implementation details

Following Figure 1, the input documents are first normalized with the pre-processing module from textacy[b] and parsed with the dependency parser in *spaCy*. The identification model and the ranking module are constructed with modules from *PyTorch*, and trained with the training datasets mentioned above. Term clustering is performed in the following procedure: i) use sentence-transformers to transform ngrams to sentence vectors; ii) use the module of agglomerative clustering in *sk-learn* to cluster the sentence vectors to groups. The TF-IDF data is obtained from all the parsed documents of all the ten datasets.

### 4.2 Evaluation of candidate Identification

The candidate identification component is aimed at identifying as many candidate terms as possible among the ngrams generated from a document. With the assumption that non-keywords should have low keyness scores and should be dealt with in the candidate ranking component, only the metric of recall is used to evaluate the performance of candidate identification, i.e.,

$$R_{identification} = \frac{Number\ of\ correctly\ identifiied\ keywords\ in\ a\ document}{Number\ of\ gold\ keywords\ in\ a\ document} \qquad (12)$$

Following the training procedure in Section 3.4, all the instances in the training data of the ten datasets are used for training. The number of training epochs is 20, batch size is 56, learning rate is 1, and the filter $\varepsilon$ is set to be one of [0.5, 0.75, 1]. The recall scores are given in Figure 4. The average recall score for $\varepsilon = 0.5$ is 0.958, the average score for $\varepsilon = 0.75$ is 0.951, and the average score for $\varepsilon = 1.0$ is 0.936. The data show that there is negative correlation between the filter $\varepsilon$ and the recall score. The increase of $\varepsilon$ means more positive unlabelled data are used in the training process, leading to decrease of recall score in the identification process.

---

[b]The Python packages and their versions used in this study are listed below: *textacy* (0.13.0), *spaCy* (3.7.2), *sklearn* (1.3.0), *PyTorch* (2.1.1+cu121), and *sentence-transformers* (0.2.1). The python scripts of PKE will be released to the public via github (https://github.com/) when ready.

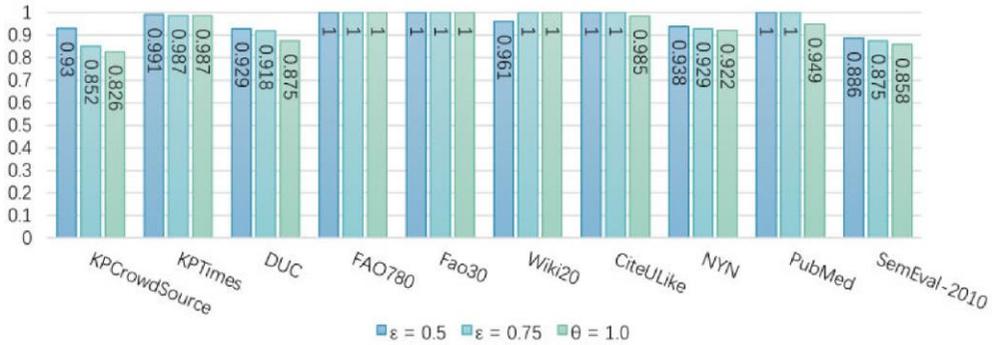

**Figure 4.** Performance of candidate identification on the ten datasets.

Figure 4 shows that CNN-based keyword identification is advantageous over heuristic part-of-speech-based rules. Regular expression templates like *(ADJECTIVE)\*(NOUN)+* are used in Wan and Xiao (2008), Ushio et al. (2021), Florescu and Caragea (2017), Boudin (2018), and Hassani et al. (2022) etc. Although researchers observe that the vast majority of keywords follow such templates (Ushio et al. 2021; Justeson and Katz 1995), the statistics in Table 4, which are derived from the ten datasets used in this study, shows that better performance can be obtained with supervised methods. In the table 87.33% keywords end with nouns, 6.14% with verbs, and 2.08% with adjectives, and the rest of 4.45% keywords end with other parts of speech. Using the above regular expression template should be able to recall 87.33% of keywords, which is far lower than the best performance of CNN-based model (0.965).

**Table 5.** Statistics of keywords ending with particular parts of speech

| Ending with | Percentage | Number of patterns | Example |
|---|---|---|---|
| Noun | 87.33 | 1997 | JJ NN: computational intelligence<br>NN NN: film producer<br>NNP VBD NN: oscar nominated actress<br>JJ NN NN IN JJ NNS: original soap house of dark shadows<br>VBG DT NN IN DT NN IN DT NN:seeking a friend for the end of the world |
| Verb (-ING and -ED, etc.) | 6.14 | 267 | JJ NN VBG: consistent query answering<br>NN HYPH VBN: Object-based<br>NNP VBZ: Senate approves |
| Adjective | 2.08 | 88 | VB JJ: inject malicious<br>JJ JJ: derivative free<br>RB JJ: blatantly unconstitutional |

NN stands for the part of speech NOUN, JJ for ADJ, VBG for -ING verb form, VBN for past participle, VBZ for third singular form, RB for adverb, VBD for -ED verb form, DT for determiner.

### 4.3 Evaluation of candidate ranking

#### 4.3.1 Evaluation metrics for candidate ranking

Two metrics are used to evaluate PKE's ranking performance. One is the top-10 F-measure computed on the basis of top-10 precision and top-10 recall, which is the most traditional and popular measure used for keyword extraction task (Firoozeh et al. 2020), given below:

$$P_{top10} = \frac{\text{Number of correctly extracted keywords in a document}}{10} \qquad (13)$$

$$R_{top10} = \frac{Number\ of\ correctly\ extracted\ keywords\ in\ document}{Number\ of\ assigned\ keywords\ in\ document} \tag{14}$$

$$F_{top10} = \frac{2 \times P_{top10} \times R_{top10}}{P_{top10} + R_{top10}} \tag{15}$$

As PKE outputs ranked term groups (ref. Figure (1)), what $F_{top10}$ measures are not terms, but term groups (or topics). A term group is considered correct if any term inside the group matches the gold keywords assigned to the document in question.

The second evaluation metric is mean reciprocal rank (MRR) (Papagiannopoulou and Tsoumakas 2020), computed as below:

$$MRR = \frac{1}{|P|} \sum_{t \in P} \frac{1}{rank_t} \tag{16}$$

wherein $t$ is a term identifiable in the gold answer in a given document, $P$ is the list of predicted keywords, $rank_t$ is the rank of $t$ in $P$. MRR measures the average of the reciprocal ranks of the first correct keyword in the predicted list. As the measure is associated with a user model where the user only wishes to see one relevant item (Craswell 2009), a higher MRR score indicates that the model is more capable of retrieving keywords that are closer to human interpretation of the documents.

### 4.3.2 Performance comparison with state-of-the-art systems

Table 5 lists the performance of PKE and three state-of-the-art systems including RaKUn (Škrlj et al. 2019), YAKE (Campos et al. 2020), and TNT-KID (Martinc et al. 2022), among which RaKUn and YAKE are unsupervised systems and TNT-KID is a supervised system. The PKE ranking model is trained with all the training data from the ten datasets in 30 epochs, with batch size set to be 126, learning rate 0.0008, and the sampling ratio $\vartheta$ 3.35. The table shows that in comparison with the three systems PKE achieves the best F-scores in seven of the ten datasets, comparable F-scores in one dataset, and lower F-scores in the other two datasets. In particular, the F-score that PKE obtains in the dataset SemEval-2010 is 36.7%, higher than TNT-KID. When summarizing the workshop on SemEval-10, Kim et al. (2013) comment that an F-score of 100% for top-k F-measure is infeasible because human readers can only achieve an F-score of 33.6% against author-assigned keywords. The F-score obtained by PKE demonstrates that PKE is capable of performing keyword extraction as good as human readers and is more consistent than human readers.

**Table 6.** Top10 F-scores of PKE and three state-of-the-art systems. PKE scores are obtained with $\vartheta = 3.35$

| Name | Metric | KP-Crowd-souce | KP-Times | NYN | SemEval-2010 | PubMed | DUC | FAO-380 | FAO-30 | Wiki-20 | CiteU-Like |
|------|--------|--------|--------|------|--------|--------|------|---------|--------|---------|-----------|
| unsupervised | | | | | | | | | | | |
| RaKUn | F-score | 0.428 | — | 0.096 | 0.091 | 0.075 | — | 0.094 | 0.233 | 0.190 | — |
| YAKE | F-score | 0.173 | — | 0.256 | 0.211 | 0.106 | — | 0.187 | 0.184 | 0.162 | 0.256 |
| supervised | | | | | | | | | | | |
| TNT-KID | F-score | — | 0.485 | 0.358 | 0.337 | — | 0.373 | — | — | — | — |
| PKE | F-score | 0.334 | 0.201 | 0.335 | 0.367 | 0.220 | 0.404 | 0.278 | 0.363 | 0.328 | 0.329 |
| PKE | MRR | 0.353 | 0.262 | 0.410 | 0.356 | 0.348 | 0.411 | 0.350 | 0.404 | 0.370 | 0.340 |

In comparison with RaKUn, PKE achieves better F-scores on shared datasets except for KP-Crowdsource. Using load centrality as the ranking feature, RaKUn obtains a much higher F-score (0.428) on KP-Crowdsource than PKE, but a comparatively lower F-scores (0.233) on FAO30, and very much lower F-scores on NUS, PubMed, and SemEval2010 (respectively 0.096, 0.075, and 0.091). This might be due to the fact that load centrality is a significantly contrastive keyness feature in datasets like KP-Crowdsource, but is insignificant in other datasets. Load centrality is a kind of betweenness centrality of a vertex inside a graph computed on the number of shortest paths that pass through the vertex Goh et al. (2001); Škrlj et al. (2019). As shown in Figure 5 and Figure 6, the two betweenness centrality values of keywords in KP-Crowdsource ranges from 0 to 1, while the range of keywords in NYN is [0, 0.28], a much narrower range. Furthermore, the range of non-keywords in NYN possess a much wider range ([0, 0.37]) than keywords in the dataset. As the texts in KP-Crowdsource are mainly news items (Marujo et al. 2012) and NYN dataset consists mainly of scientific papers, it is speculated such contrast in betweenness centrality is due to differences between the two sublanguage domains.

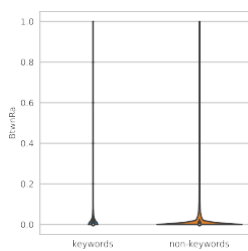

**Figure 5.** Betweenness centrality scores of KP-Crowdsource

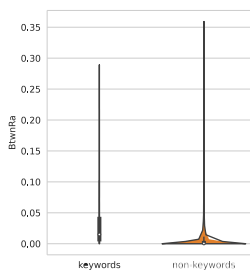

**Figure 6.** Betweenness centrality scores of NYN

In Table 6, the F-scores of PKE are higher than YAKE. This is mainly attributed to the reason that PKE is a supervised approach while YAKE is unsupervised, although PKE and YAKE share six features (with different calculating methods for casing, position, frequency, and TF-IDF) in the ranking process. Figure 7 gives the contrastive violin plots of the six features of keywords (88,012 instances) and non-keywords (10,888,790 instances), showing that the six features are effective keyness indicators despite the disturbance from positive unlabelled data. Another reason that PKE better performs is that it uses more information (a total of 19 features), while YAKE mainly uses heuristic features.

In comparison with TNT-KID, PKE achieves higher F-scores on DUC and SemEval-2010, but lower scores on KP-Times and NYN. Although both TNT-KID and PKE use supervised approach, they formulate keyword extraction as different tasks. TNT-KID treats keyword extraction as a sequence-labelling task and trains a transformer encoder in two phases: language model pretraining on a large corpus and fine tuning for keyword identification with keyword datasets(Martinc et al. 2022). The performance of fine-tuning systems is expected to be better when large amount of data are available. This explains why TNT-KID achieves comparatively higher F-scores on KP-Times (with 259,923 documents as training data). Nevertheless, systems like TNT-KID are domain dependent, even though pre-training might help reduce to a certain degree the reliance on training data size. PKE, on the other hand, treats keyword extraction as pattern-based ranking learning task. The purpose of training is to learn to rank a pattern according to known keyness patterns. Given the assumption that the number of keyness patterns is limited (ref. Section 4.3.3), PKE at its present form may not always benefit from the increase of dataset size. This is because the number of positive unlabelled instances should increase as the dataset size increases, leading to the increase of keyness patterns that are labelled as non-keywords, while the increase of dataset size does not add to the number of labelled keyness patterns. Therefore, PKE prefers datasets that are more consistent in keyword assignment, but not bigger datasets. This explains why PKE obtains a lower F-score on KP-Times, but higher F-scores on DUC and SemEval-2010.

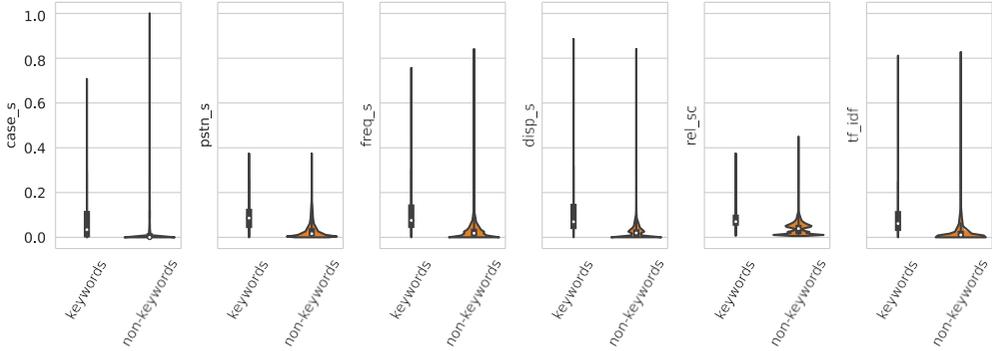

**Figure 7.** Shared features between PKE and YAKE. *Case_s* stands for casing score, *stn_s* position score, *freq_s* for frequency score, disp_s for dispersion score, rel_sc relative context score, and *tf-idf* for TF-IDF score.

We argue that PKE achieves state-of-the-art performance with two rationales. First, the three systems selected for comparison in Table 6 achieve comparatively higher F-scores at the time we survey the literature. Second, researchers have suggested that it is not possible to achieve very high F-scores in the task of keyword extraction. Turney (2000) suggests that an ideal keyword extraction algorithm could in principle recall about 75% of the human-generated keywords. As the agreement between subjects was around 60% in experiments, the highest score for an algorithm based on learning from human judgment might be about 60%-75% in agreement with the gold answers (Hjørland 2001).

### 4.3.3 Experiments on cross-domain keyword extraction

This section reports zero-shot experiments conducted to evaluate PKE's performance in cross-domain keyword extraction. The experiments are zero-shot as we exclude from the training process the training data of four datasets, i.e., DUC of miscellaneous news, FAO30 of articles on agriculture, Wiki20 of papers of computer science, and SemEval-2010 of miscellaneous papers. The obtained model is then used for keyword extraction on the test data of the four datasets. The experiment results are given in Table 6.

In comparison with the F-measures of the four dataset in Table 6, The F-score of DUC in Table 7 increases by 0.26, while Wiki-20 drops by 0.027, FAO30 drops by 0.032, and SemEval-2010 drops by 0.044. PKE's performance on the four datasets do not deteriorate even though their training data are not included in the training process. The system is capable of providing state-of-the-art keyword extraction performance for unseen data, given only the specification of sublanguage type, such as the "miscellaneous news" for DUC and "articles on agriculture" for FAO30. We attribute this capacity to the posit that keyness patterns at the community level are to a certain extent universal across language domains, the mechanism designed in Equation (1) that is implemented by the ranking model in Figure 3.

**Table 7.** Cross-domain performance of PKE on four datasets

| Metric | DUC | Wiki-20 | FAO30 | SemEval-2010 |
|---|---|---|---|---|
| F-measure@10 | 0.430 | 0.301 | 0.331 | 0.323 |
| Precision@10 | 0.38 | 0.400 | 0.567 | 0.393 |
| Recall@10 | 0.512 | 0.244 | 0.236 | 0.281 |
| MMR@10 | 0.332 | 0.395 | 0.354 | 0.376 |

*Keyness patterns and training data size*

PKE's capacity in cross-domain keyword extraction relates to the community-shared keyness patterns. Although it is a general rule that supervised machine learning are sensitive to input training data size, but the relationship between the performance and training data size in PKE is not linear. PKE also requires a large amount of data for training, but the increase of training data size does not lead to performance increase.

The above observation is demonstrated in Figure 8, which plots the non-linear relationship between the number of instances and the accumulative percentage of pattern types using the following procedure: (i) Cluster the dependent feature vectors of all the keywords (88,011 instances) in the training data of the ten datasets by setting the cosine distance threshold to be 0.1 and obtain 297 clusters; (ii) Compute the accumulative percentage of clusters that the number of instances from 1 to 88,011. As the cosine distance threshold is 0.1, the feature vectors in each cluster should be fairly similar (with cosine similarity bigger than 0.9). As the figure shows, about 60% clusters are covered in the first 2,000 instances and 85% clusters covered in about 20,000 instances. The cluster coverage is almost 100% when the number of instances reaches 60,000. Therefore, it can be inferred that the increase of input training data may not lead to better performance when the size of input training data reaches 60,000.

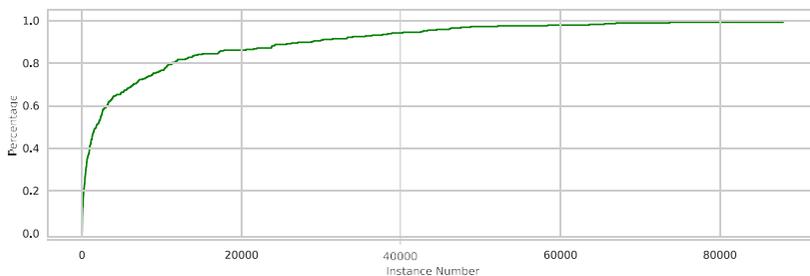

**Figure 8.** Non-linearity between instance number and accumulative percentage of pattern types

Note that the statistics in Figure 8 encompass all sublanguage domains. That is, the delinearity between training data size and keyword extraction performance is not sensitive to sublanguage domain. This explains why PKE is robust in performing zero-shot cross-domain keyword extraction. The keyness patterns learned in other language domains should suffice for state-of-the-art performance.

*Independent features vs. dependent features*

The distinction between independent features and dependent features also contributes to PKE's performance. The impact of language domain on dependent features is illustrated by the comparison between Figure 9 and Figure 10, which respectively give the dependent feature statistics of DUC and MedPub. DUC consists of texts of news, whereas MedPub of papers on medicine. The first obvious contrast between the two datasets is that the scales of some features of DUC are generally higher than those of MedPub except for lexical specificity and relative entropy. Some feature contrasts can be easily explained by the differences between news texts and academic papers of medicine, such as position scores and betweenness centrality. News articles generally are shorter but more diverse in topics. Therefore, the keywords of news articles are more likely characterized with position scores and higher betweenness centrality values, whereas academic papers generally are longer but more coherent and more focused and keywords inside these papers may have lower position scores and lower betweenness centrality values. The impact of term length on dependent features is also observed in the experiments. In particular, the differences between bigrams and quagrams are very obvious.

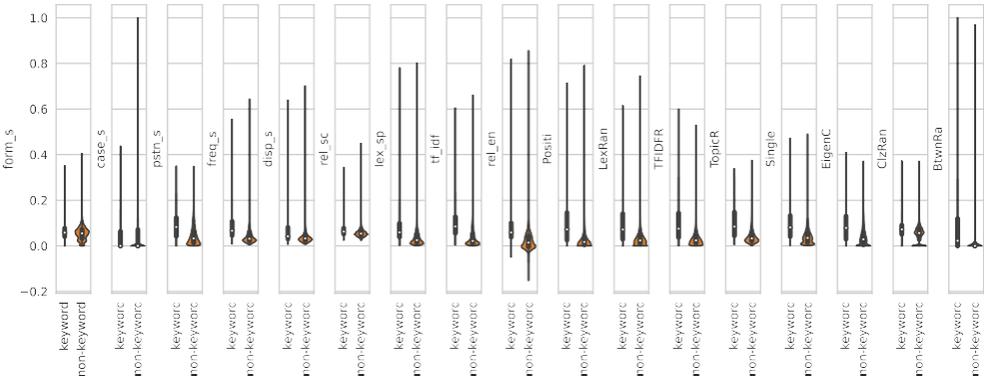

**Figure 9.** Violin plots of dependent features of DUC

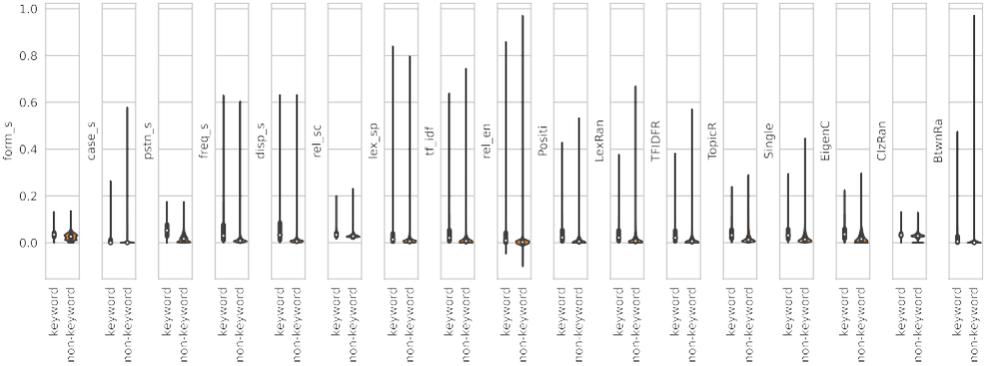

**Figure 10.** Violin plots of dependent features of MedPub

The F-scores in Table 7 show that the mechanism used in PKE is effective in dealing with feature pattern contrasts introduced by language domains and term length. With the embeddings of sublanguage type and term length learned via the training process, the dot product introduced in Figure 3 is capable of adjusting the input dependent feature vectors so that pattern complexity is reduced before transformer encoders and convlutional transformation are applied for pattern recognition.

### 4.3.4 Experiments with Positive Unlabelled Learning

As discussed in Section 3.4, PKE is trained as a task of Positive Unlabelled Learning and there exists a sampling ratio value $\vartheta$ that balances both excess risk and the number of negative training data so that the best performance could be achieved. This observation is supported by the experiments with the ten datasets. Figure 11 depicts how the sampling ratio $\vartheta$ affects the ranking performance. In the figure, the x axis is $\vartheta$ within the range [0.1, 4.1]. The increase of the sampling ratio indicates the increase of sample size, but the contamination ratio $\hat{\sigma}$ remains unchanged. The y axes are respectively the average recall, average precision, and average F-measure of the 10 datasets. Linear regression applied on the three average scores shows that the increase of sample size should increase the excess risk of discriminating P when both the size of positive labelled data (P) and the contamination ration $\hat{\sigma}$ are fixed, leading to the decrease of overall performance. But it also allows for more negative instances to participate in the training process. According to the plots in the figure, PKE achieves maximum F-scores when $\vartheta$ is in the range [1.6, 3.6].

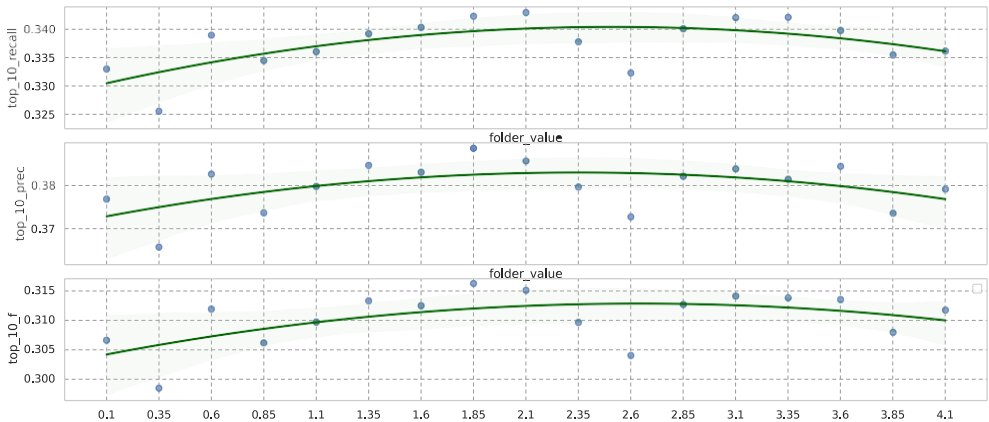

**Figure 11.** Correlation between folder ratio and ranking performance. Solid lines are linear regression results.

The relation between the rate of annotation agreement and contamination ratio $\hat{\sigma}$ can also be observed by examining PKE's performance on individual datasets. Less consistency ratio leads to higher contamination ratios. A dataset with less consistency ratio should require a higher $\vartheta$ to reach the balance for the best performance. This phenomenon is observed by the comparison between Figure 12 and Figure 13, which respectively depicts PKE's performance on DUC and KP-Crowdsource. According to Wan and Xiao (2008), DUC consists of 308 standard news which are manually labelled with at most 10 key-phrases and the Kappa statistic for inter-agreement among annotators is 0.70. Nevertheless, KP-Crowdsourcing is annotated via crowdsourcing and the average agreement between annotators is 55% (Marujo et al. 2012). The $\vartheta$ for DUC should be lower than that of KP-crowdsource. Accordingly, DUC reaches balance at around 1.35 (ref. Figure 12) and the balance value of $\vartheta$ for KP-Crowdsource is beyond 4.1 (ref. Figure 13).

The impact of the sampling ratio $\vartheta$ on the metric of MRR differs from that of F-measure, as is illustrated in Figure 14. Within the range of [0.1, 4.1], MRR decreases along with $\vartheta$. The decrease of MRR means that the overall ranking positions of the identified keywords are moving away from the top. This phenomenon can be attributed to the increase of unlabelled positive instances in the training dataset. The increase of unlabelled positive instances is expected to have an impact on those labelled keyness patterns, leading to the decrease of MRR.

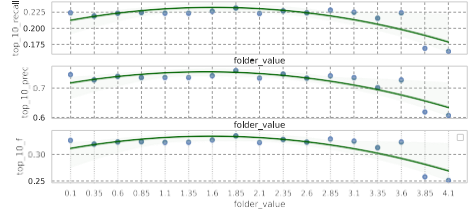

**Figure 12.** PKE's performance on DUC with different contamination ratio

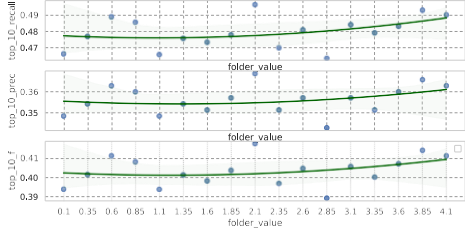

**Figure 13.** PKE's performance on KP-crowdsource with different contamination ratio

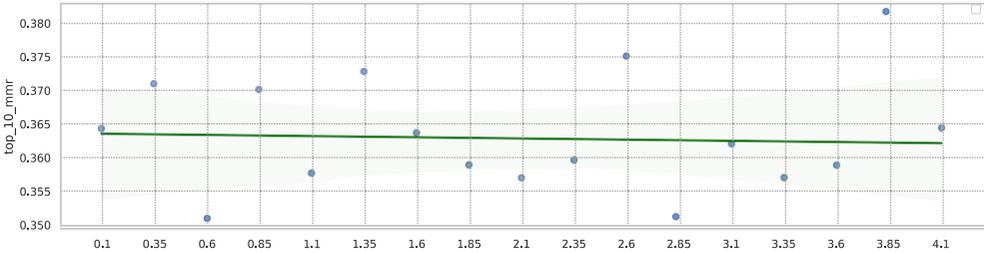

**Figure 14.** The impact of sampling ratio on MRR

## 5. Conclusions

There has been no consensus among researchers on the essence of supervised keyword extraction and on how the task should be formulated. With the posit that human annotators generally utilize keyness patterns when deciding on keywords of a document, this study formally defines keyness patterns by distinguishing independent features and dependent features, and proposes a CNN-based framework to imitate the keyword assignment behaviour performed by human beings. Experiments with ten datasets demonstrate that PKE is advantageous in the following aspects:

(1) As a supervised keyword extraction approach, PKE obtains state-of-the-art F-scores on small, middle, and large keyword datasets;

(2) Zero-shot experiments demonstrate that PKE is very robust in cross-domain keyword extraction and is requisite of no training data of particular datasets.

The state-of-the-art performance of PKE is attributed to the fact that keyness patterns are generalizable across different keyword datasets and that keyness patterns that people use in annotating keywords are not diverse but are limited in number. In addition, experiments with PKE also reveal that supervised keyword extraction should be treated as a positive unlabelled learning task because subjectivity is intrinsic in human beings' behaviour of keyword assignment. There exists a sampling size for unlabelled data that balances both excess risk and the size of negative training data so that the best performance can be achieved.

Although the experiments prove that pattern-based keyword extraction is a close approximation of human beings' behaviour of keyword assignment, many aspects of the proposed approach await further exploration. For instance, this study identifies two independent features and seventeen dependent features. But further research is required to understand the nature of these features and to find out whether there are more effective features available for use. In addition, there might be more effective neural network models for candidate identification and candidate ranking. More importantly, this study has used only the datasets of English language. Tests should be conducted on other languages to examine whether there are general keyness patterns among different languages.